\documentclass{PoS}

\usepackage[T1]{fontenc}  
\usepackage{amsmath}
\usepackage{epstopdf}
\usepackage{graphicx}
\usepackage{floatrow}

\newcommand{\nl}{n_{\ell}}

\newcommand{\mbar}{\overline{m}}
\newcommand{\MSR}{\mathrm{MSR}}

\newcommand{\MSb}{\overline{\mathrm{MS}}}
\newcommand{\df}{{\rm d}}

\newcommand{\LQCD}{\Lambda_\mathrm{QCD}}

\title{Cornell Model calibration with NRQCD at N$^3$LO}

\ShortTitle{Cornell Model calibration with NRQCD at N$^3$LO}

\author{\speaker{Pablo G. Ortega}$\,\,\textsuperscript{,1,}$
\thanks{This work has been partially funded by the Spanish MINECO {\it Ram\'on y 
 Cajal 
program} (RYC-2014-16022), the MECD grants FPA2016-78645-P, FPA2016-77177-C2-2-P 
and the IFT {\it Centro de 
Excelencia Severo Ochoa} Program under Grant SEV-2012-0249. P.G.O. acknowledges 
the financial support from {\it Junta de  Castilla y Le\'on} and European 
Regional Development Funds (ERDF) under Contract 
no. SA041U16, and by Spanish MINECO's {\it Juan de la Cierva-Incorporaci\'on} 
program with grant agreement no.~IJCI-2016-28525.}\\
        E-mail: \email{pgortega@usal.es}}
        
\author{Vicent Mateu$\,\,\textsuperscript{1,\,2}$\\
        E-mail: \email{vmateu@usal.es}}

\author{David R. Entem$\,\,\textsuperscript{1}$\\
        E-mail: \email{entem@usal.es}}

\author{Francisco Fern\'andez$\,\,\textsuperscript{1}$\\
	E-mail: \email{fdz@usal.es}\\ \vspace*{0.5ex}

	$\textsuperscript{\bf 1}$  Departamento de F\'isica Fundamental and IUFFyM, Universidad de Salamanca\\
	Plaza de la Merced S/N, E-37008 Salamanca, Spain\\
	$\textsuperscript{\bf 2}$  Instituto de F\'isica Te\'orica UAM-CSIC\\C/ Nicol\'as Cabrera 13-15,
	Campus de Cantoblanco, E-28049 Madrid, Spain\\}

\abstract{The typical binding energy of heavy hadron spectroscopy makes the system 
accessible to perturbative calculations in terms of non-relativistic QCD. Within 
NRQCD the predictions of heavy quarkonium energy levels rely on the accurate 
description of the static QCD potential $V_{\rm QCD}(r)$.

Historically, heavy quarkonium spectroscopy was studied using phenomenological 
approaches such as the Cornell model $V_{\rm Cornell}=-\kappa/r+\sigma\, r$, which 
assumes a short-distance dominant Coulomb potential plus a liner rising
potential that emerges at long distances. Such model works reasonably well in 
describing the charmonium and bottomonium spectroscopy. However, even when there 
are physically-motivated arguments for the construction of the Cornell model, 
there is no conection a priori  
with QCD parameters.

Based on a previous work on heavy meson spectroscopy, we calibrate the Cornell 
model with NRQCD predictions for the lowest lying 
bottomonium states at N$^3$LO, in which the bottom mass is varied within a wide 
range. We show that the Cornell model mass parameter can be identified with the 
low-scale short-distance  MSR mass at the scale $R = 1$ GeV. This identification 
holds for any value of $\alpha_s$ or the bottom mass. For moderate values of $r$, 
the NRQCD and Cornell static potentials are in head-on agreement when switching 
the pole mass to the MSR scheme, which allows 
to simultaneously cancel the renormalon and sum up large logarithms.
}

\FullConference{XIII Quark Confinement and the Hadron Spectrum - Confinement2018\\
		31 July - 6 August 2018\\
		Maynooth University, Ireland}

\begin{document}

\section{Introduction}

\noindent
The so-called November Revolution in 1974~\cite{Augustin:1974xw,Aubert:1974js} 
triggered numerous theoretical developments in hadron spectroscopy. 
Non-relativistic phenomenological models were justified due to the large mass of 
the charm and bottom quarks inside mesons, and counterbalanced the lack of 
development of Quantum Chromodynamics (QCD) for heavy quarkonium systems at that 
time. 
Among those early studies we highlight the works of 
Eichten~\cite{Eichten:1978tg}, Godfrey~\cite{Godfrey:1985xj}, 
Stanley~\cite{Stanley:1980zm} or Bhanot~\cite{Bhanot:1978mj}, which employed an 
unified and simple framework to phenomenologically study both light- and 
heavy-meson spectroscopy.

Those aforementioned phenomenological models considered quarks as low-energy 
spectators, which interact through
flavor-independent gluonic degrees of freedom. The short-distance interaction is 
assumed to be of perturbative nature, dominated by a single t-channel gluon 
exchange (that is, a Coulomb interaction proportional to $\alpha_s$, the strong 
coupling constant at some scale). At long distances, non-perturbative effects are 
expected to emerge, and are modelled with a linear confining interaction, 
confirmed by lattice QCD calculations~\cite{Bali:2000vr}.
Hence, the simplest quark model for heavy quarkonium is the so-called {\it 
Cornell potential}\,:
\begin{align}\label{eq:Cornellpot}
V_{\rm Cor}(r)=\sigma\,r\,-\,\frac{C_F\,\alpha_s^{\rm Cornell}}{r}\,.
\end{align}
Here $\alpha_s^{\rm Cornell}$ and $\sigma$ are purely phenomenological constants 
of the model with, at first glance,
no direct connection with QCD parameters. $C_F$ is the first $SU(N_c)$ Casimir 
[\,$C_F= (N_c^2-1)/(2N_c) =4/3$ for $N_c=3$\,].

Despite of its simplicity and ``ad hoc'' construction, such model is able to 
reproduce the heavy quarkonium spectra, which suggests that a relation, even 
vague, may exist between the Cornell model and the more theoretically-solid 
non-relativistic QCD (NRQCD).
NRQCD~\cite{Lepage:1987gg} is obtained from QCD by integrating out the heavy 
quark mass $m_Q$. A perturbative matching can be performed exploiting that 
$m_Q\gg\Lambda_{\rm QCD}$, so NRQCD inherits all the light degrees of freedom from QCD. 
However, NRQCD mixes soft and ultrasoft scales, complicating the power-counting. 
Solutions to this problem appeared with the development of different 
EFTs such as velocity NRQCD (vNRQCD)~\cite{Luke:1999kz} and
potential NRQCD (pNRQCD)~\cite{Pineda:1997bj,Brambilla:1999xf}, which describe 
the interactions of a non-relativistic
system with ultrasoft gluons, systematically organizing the perturbative 
expansions in $\alpha_s$ and the velocity of heavy quarks. Such EFTs only 
include the relevant degrees of freedom for $Q\overline Q$ systems near 
threshold, integrating out the rest of degrees of freedom.

In this work we explore the connection between the simplest realization of the 
Cornell model against NRQCD. 
To that end we compare the mass of the lowest-lying $Q\overline{Q}$ bound 
states, observables that can be reliably predicted both in the theory and the 
model, varying the quark mass and the strong coupling constant. We show that the 
Cornell potential agrees for large values of $r$ with the QCD static potential 
once the latter is expressed in terms of the MSR mass and improved with 
all-order resummation of large renormalon-related logs via R-evolution~\cite{Hoang:2017suc}. 
Our R-improved static potential also compares nicely with lattice QCD simulations 
from Refs.~\cite{Bazavov:2014pvz,Bazavov:2017dsy}.

A more complete description of the methods and further results can be found in Ref.~\cite{Mateu:2018zym}.

\section{Cornell Potential and QCD Static Potential for $\mathbf{Q\overline Q}$}\label{sec:StaticPot}

\noindent
As mentioned above, our main goal is to compare the Cornell potential predictions with NRQCD, 
so relations among QCD fundamental constants and Cornell model parameters can be obtained.

On the one hand, the Schr\"odinger equation for the Cornell potential of Eq.~\eqref{eq:Cornellpot} is 
solved numerically using the Numerov algorithm~\cite{Numerov1}, from where we calculate the mass
of the bound states. However, such potential is not sufficient to predict the  $\Upsilon(1S)-\eta_b(1S)$
or  $\chi_{bJ}$ multiplet (with $J=\{0,1,2\}$) mass splittings. Further spin dependence is needed in the
Cornell static potential. For that reason, it is important to add $1/m_Q^2$ terms to take into account the
spin-spin, spin-orbit and tensor interactions, breaking the leading power degeneracy~\cite{DeRujula:1975qlm}.
Such contributions emerge from the s-channel gluon exchange and the leading relativistic 
corrections of the t-channel gluon and confinement interactions~\cite{Godfrey:1985xj}\,:
\begin{align}
V_{\rm SS}^{\rm OGE}(r)&=\frac{8\alpha_s^{\rm Cornell}}{9m_Q^2\,r^2}(\vec S_1\cdot \vec S_2)\,\delta(r)\,,\label{eq:subleading}
&V_{\rm LS}^{\rm OGE}(r)&=\frac{2\alpha_s^{\rm Cornell}}{m_Q^2\,r^3}\,(\vec L\cdot\vec S)\,,\\
V_{\rm LS}^{\rm CON}(r)&=-\frac{\sigma}{2m_Q^2\,r}\,(\vec L\cdot \vec S)\,,
&V_{\rm T}^{\rm OGE}(r)&=\frac{\alpha_s^{\rm Cornell}}{3m_Q^2\,r^3}\,S_{12}\,,\nonumber
\end{align}
being $\vec S=\vec S_1+\vec S_2$ the total spin, $\vec L$ the relative orbital momentum and $S_{12}$ the
tensor operator of the $Q\overline Q$ bound state, defined as 
$S_{12}=2\,(\vec S_1\cdot\hat r)(\vec S_2\cdot\hat r)-(\vec S_1\cdot\vec S_2)$, with $\hat r= {\vec r}/r$.
With no significant loss of precision, such contributions can be calculated 
using first-order perturbation theory, given the large mass of the heavy quark.

Using the above potential we are able to fit the parameters to the low-lying experimental bottomonium spectrum, those with $n_p=\{1,2\}$, whose masses are taken from the PDG~\cite{Tanabashi:2018oca}. Such fit gives the values\,:
\begin{align}\label{eq:Cornell-Fit}
\{m_b^{\rm Cornell}\,,\,\sigma\,,\,\alpha_s^{\rm Cornell}\} & = 
\{4.733 \pm 0.018\,{\rm GeV}\,,\, 0.207 \pm 0.011\,{\rm GeV}^2\,,\, 0.356 \pm 0.015\,\}
\end{align}
where the uncertainties correspond to the $68\,\%$ confidence level for each parameter. Given the extremely precise nature of the experimental masses, the uncertainties on the parameters that come out of the fit are penalized, increasing their values by the square root of the reduced $\chi^2$ at its minimum.

On the other hand, within NRQCD we can define the static potential as the color-neutral interaction
between two infinitely heavy color-triplet states. Such potential has been calculated up to $\mathcal{O}(\alpha_s^4)$,
at which point the potential becomes time-dependent and the static approximation breaks down.
Such feature derives in a dependence on the ``ultrasoft'' scale $\mu_{\rm us}$.

In position space and taking the charm quark as massless, the perturbative contribution to the static QCD potential can
be written as follows\,:
\begin{equation}\label{eq:static}
V^{(\nl)}_{\rm QCD}(r,\mu) = -\,C_F\,\frac{\alpha^{(\nl)}_s(\mu)}{r}\sum_{i=0}\sum_{j=0}^{i}
\biggl(\frac{\alpha^{(\nl)}_s(\mu)}{4\pi}\biggr)^{\!\!i}\, a_{i,j}(\nl)\,\log^{\,j}(r\,\mu\, e^{\gamma_E})-\,\frac{9\,C_F}{4\pi}\,\alpha_s^{(\nl)}(\mu)^4\,\frac{1}{r}\log\,(\mu_{\rm us}\,r)\,,
\end{equation}
where the coefficients $a_{i,0}$ are known to four loops~\cite{Smirnov:2009fh,Anzai:2009tm}
and $a_{i,j>0}$ can be derived from the former requiring that $V_{\rm QCD}$ does not depend on $\mu$\,:
\begin{equation}
a_{i,j}(\nl) = \frac{2}{j}\sum_{k=j}^{i}\,k\,a_{k-1,j-1}(\nl)\,\beta_{i-k}(\nl)\,.
\end{equation}

The last term in Eq.~\eqref{eq:static} depends on the ultrasoft factorization scale $\mu_{\rm us}$. Following e.g.\ Ref.~\cite{Tormo:2013tha}, we will take the following expression for the ultrasoft factorization scale\,:
\begin{align}
\mu_{\rm us}=\frac{N_c}{2}\,\mu\,\alpha^{(\nl)}_s(\mu)\,,
\end{align}
that takes into account the power counting of pNRQCD~\cite{Pineda:1997bj,Brambilla:1999xf}. 

If we plot the Cornell potential versus the QCD static potential (Fig.~\ref{fig:polebottom}) we appreciate a bad
perturbative behaviour of the latter, visible as a vertical shift of the potential in the region between $0.05\,$fm and $0.2\,$fm between different orders. Indeed, this is a clear manifestation of the $u=1/2$ renormalon of the QCD static potential, which is $r$-independent, depending only on the coefficients of the QCD beta function. 
Furthermore, no addition of further orders brings the QCD static potential closer to the Cornell model. 

It is important, thus, to cancel the static QCD potential renormalon, which exactly matches that of the pole mass except for a $-2$ factor, so that the static energy $E_{\rm stat}(r,\mu) = 2\,m_Q^{\rm pole} + V_{\rm QCD}(r,\mu)$ is renormalon free. A standard approach is to re-write the pole mass in terms of a short-distance scheme
to make the cancellation manifest. For the cancellation to take place one also needs to express the
perturbative series $\delta m_Q^{\rm SD}$ [\,Eq.~\eqref{eq:MSRdef}\,] that relates the pole mass with a short-distance mass $m_Q^{\rm SD}$ in terms of $\alpha_s(\mu)$. 
When done so, powers of $\log(\mu/R)$\,\footnote{We denote $R$ as the infrared scale of the short-distance mass.} in $\delta m_Q^{\rm SD}$ will emerge  which 
may become large if $\mu$ and $R$ are very different. Besides, from the static QCD potential definition in 
Eq.~\eqref{eq:static}, $\mu$ should depend on $r$ such that $\log(r\,\mu\, e^{\gamma_E})\sim\mathcal{O}(1)$,
therefore mass schemes
with a fixed value of $R$, such as the $\MSb$, are disfavored. 

\begin{figure}[tbh!]\centering
\floatbox[{\capbeside\thisfloatsetup{capbesideposition={right,top},capbesidewidth=6cm}}]{figure}[\FBwidth]
{\caption{Dependence of the static potential renormalization scale in Eq.~\eqref{eq:static} with $r$. In solid
blue the profile function is shown, while dashed green and dashed red show the behavior
for small ($1/r$) and large (constant) $r$, respectively. The positions in which the piece-wise function changes its
functional form are signaled by vertical dashed black lines.}\label{fig:Profile}}
{\includegraphics[width=\linewidth]{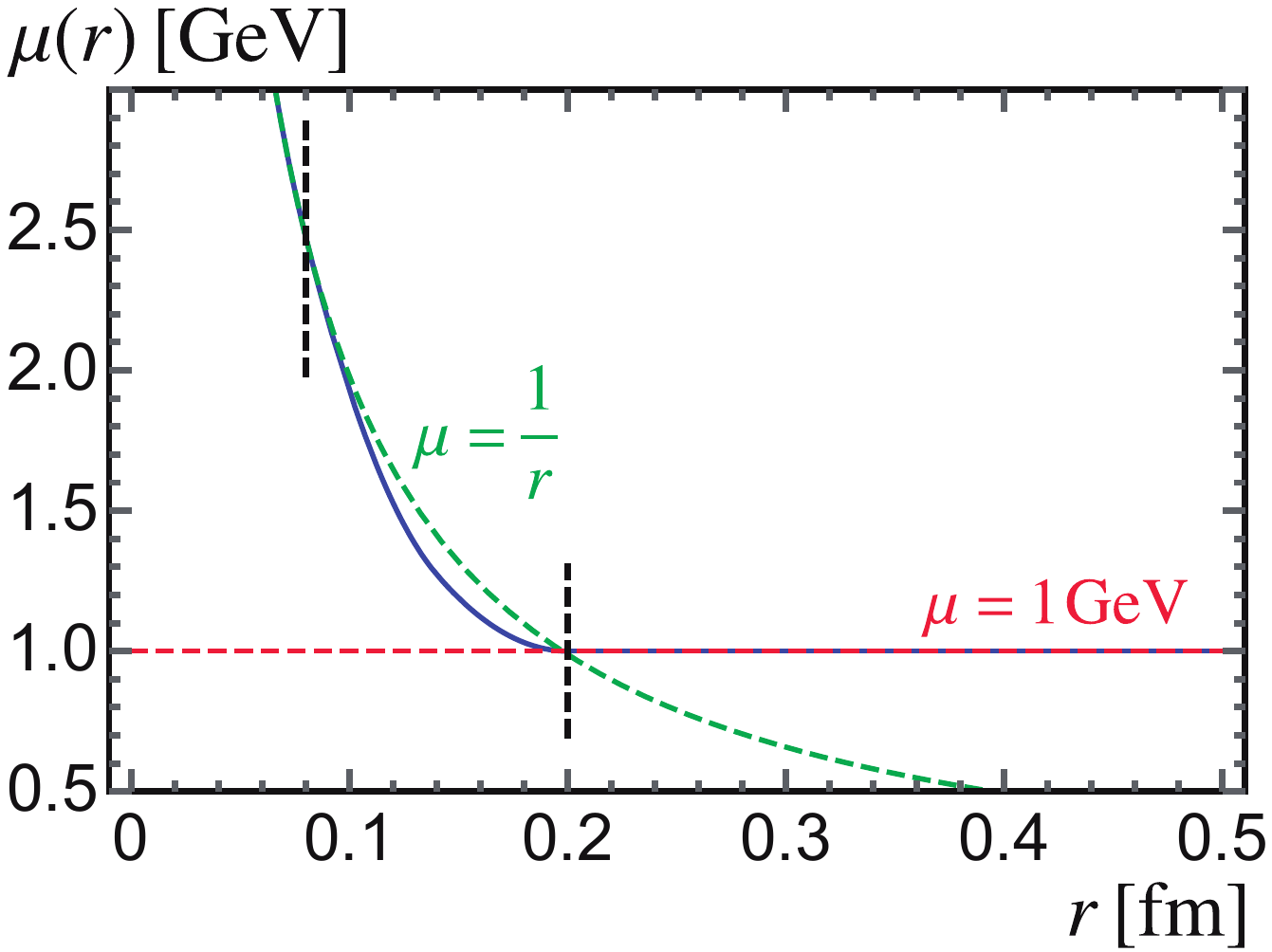}}
\end{figure}

\begin{figure*}[tbh!]
\center
{\label{fig:polebottom}\includegraphics[width=0.49\textwidth]{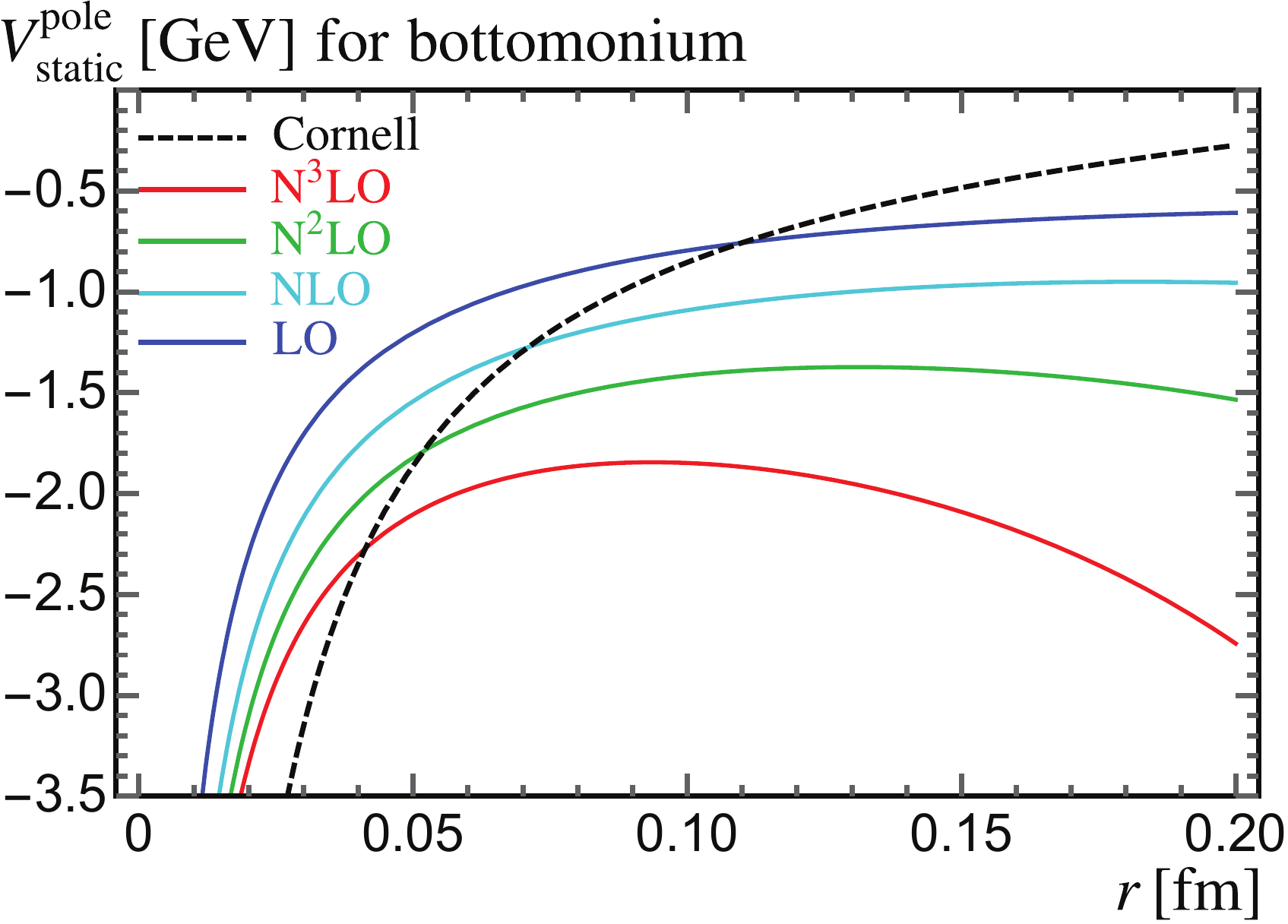}~~}
{\label{fig:MSRnbottom}\includegraphics[width=0.47\textwidth]{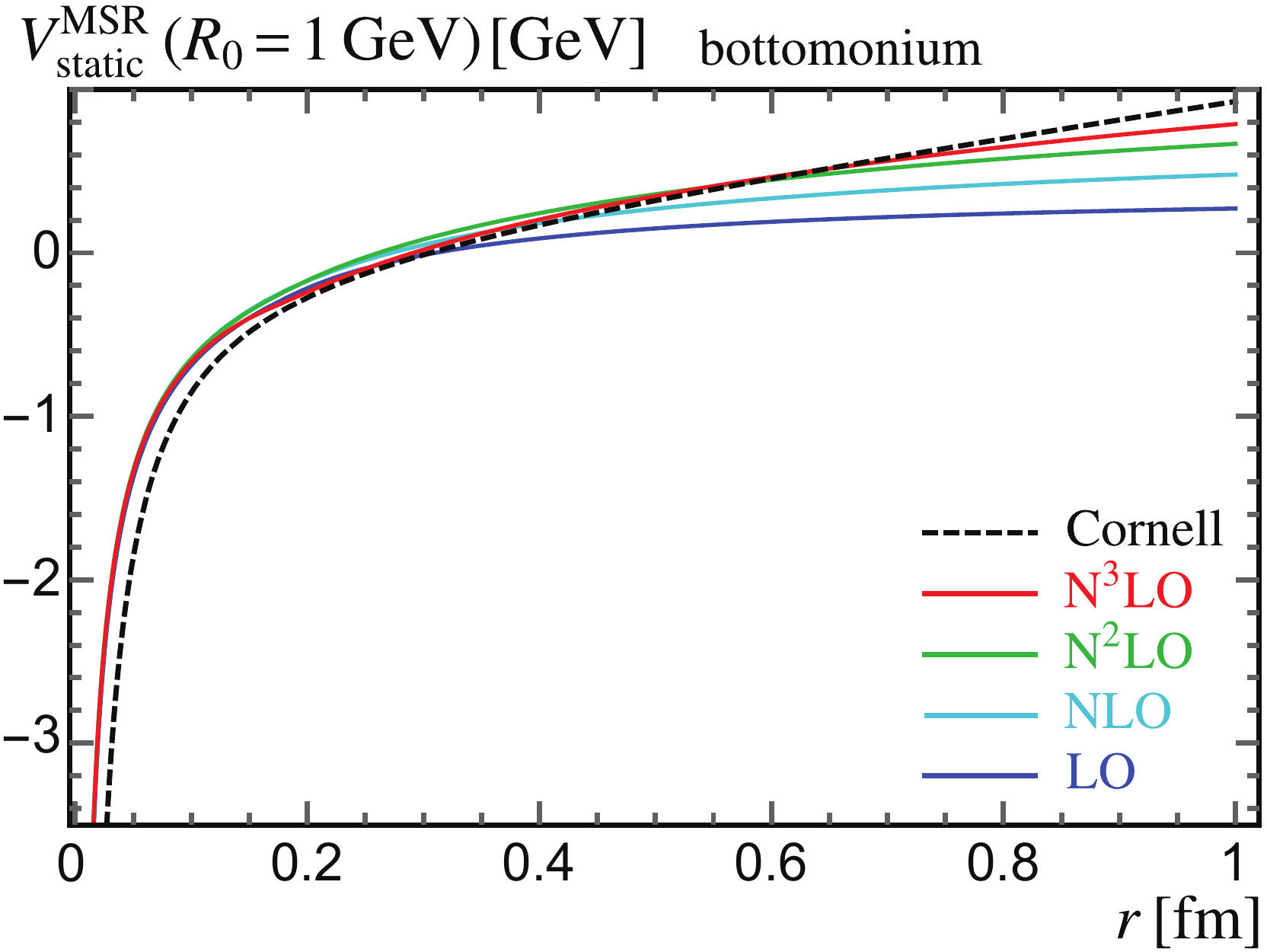} }
\caption{\label{fig:Static} Comparison of the Cornell potential (dashed black line) with the static QCD potential at
N$^3$LO (red), N$^2$LO (green), NLO (cyan) and LO (blue) for bottomonium.
The left panel shows the static QCD potential in the pole scheme, and the right panel uses the MSR scheme with the
reference scales \mbox{$R_0=1\,$GeV}. The parameters of the Cornell potential are fixed to best-reproduce the $n_p=\{1,2\}$ 
experimental bottomonium states [\,Eq.~\eqref{eq:Cornell-Fit}\,], while the Static QCD potential uses the
world average value $\alpha_s(m_Z) = 0.1181$. The left plot uses the canonical scale $\mu = 1/r$, but the right plot uses
the profile function shown in Fig.~\ref{fig:Profile}. Results for charmonium are shown in Ref.~\cite{Mateu:2018zym}.}
\end{figure*}

In order to avoid the canonical $\mu=1/r$ to rapidly acquire values in the non-perturbative region, 
we smoothly freeze it to $1\,$GeV once it reaches this value, using a transition function between $r=0.08\,$fm and $0.2\,$fm~\cite{Mateu:2018zym}. Fig.~\ref{fig:Profile} shows a graphical representation of the ``profile function'', as will be referred to.
This scale setting is similar to those implemented in the  renormalon-subtracted scheme at
Refs.~\cite{Pineda:2013lta,Peset:2018ria,Peset:2018jkf}, but with a smoother transition between the canonical and frozen regimes.

Following Ref.~\cite{Mateu:2017hlz} we use the MSR mass~\cite{Hoang:2017suc} and choose $\mu = R$ to simultaneously minimize logs in the potential and in $\delta m_Q^{\rm MSR}$.
The $\MSR$ mass scheme was created as a natural extension of the $\MSb$ mass for renormalization scales 
below the heavy quark mass. It can be directly defined from the $\MSb$-pole mass relation\,:

\begin{align}\label{eq:MSRdef}
\delta m_Q^\MSR \equiv
m_Q^{\rm pole}-m_Q^\MSR(R) & =R\sum_{n=1}^\infty \sum_{k=0}^n d_{n,k}(\nl)\biggl(\frac{\alpha^{(\nl)}_s(\mu)}{4\pi}\biggr)^{\!\!n}
\ln^k\Bigl(\frac{\mu}{R}\Bigr)\,,
\end{align}
where we see that, apart from the logarithmic dependence on the scale $\mu$, a logarithmic and linear dependence on an infrared scale R appears. The $d_{n,0}>$ are derived from the $\MSb$-pole mass relation, and depend on the specific method to change from a scheme with $(n_\ell+1)$ dynamical flavors to another with only $n_\ell$. On the one hand, the \emph{practical $\MSR$ mass} (MSRp) employs the threshold matching relations of the strong coupling to express $\alpha_s^{(n_\ell+1)}(\overline{m})$ in terms of $\alpha_s^{(n_\ell)}(\overline{m})$ and, on the other hand, the \emph{natural $\MSR$ mass} (MSRn) directly integrates out the heavy quark Q from the $\MSb$-pole relation, setting to zero all diagrams containing heavy quark loops.

The R dependence of the $\MSR$ mass is described by\,:
\begin{equation}\label{eq:Revol}
-\,\frac{\df}{\df R}m_Q^{\MSR}(R)=\gamma^R[\alpha_s^{(\nl)}(R)]=\sum_{n=0}^\infty \gamma_n^R \left(\frac{\alpha_s^{(\nl)}(R)}{4\pi}\right)^{\!\!n+1},
\end{equation}
where $\gamma_n^R(\nl) = d_{n+1,0}(\nl)-2\sum_{j=0}^{n-1} (n-j)\,\beta_j(\nl)\, d_{n-j,0}(\nl)\,$ are the R-anomalous dimension coefficients~\cite{Hoang:2017suc}, and the renormalon cancels between the first term and the sum. The solution of the RGE in Eq.~\eqref{eq:Revol} sums up
powers of $\log(R_1/R_2)$ to all orders in perturbation theory\,:
\begin{align}\label{eq:MSRsum}
\Delta^{\rm MSR}(\nl,R_1,R_2)\equiv m_Q^{\rm MSR}(R_2) - m_Q^{\rm MSR}(R_1) =
\!\int_{R_1}^{R_2}\!\df R\,\gamma_n^R[\nl,\alpha_s^{(\nl)}(R)]\,.
\end{align}
Besides, the MSR mass can be easily related to the $\MSb$ mass at the scale \mbox{$R = \mbar_Q(\mbar_Q)$}, from where it can then run down to any value of $R<\mbar_Q$.

Using the previous mass scheme we can define a short-distance potential, which is renormalon free and independent 
of the heavy quark mass. It will depend, though, on a fixed scale where the renormalon is subtracted, denoted as $R_0$.
Manipulating the static energy formula we arrive to the convenient expression\,:
\begin{align}\label{eq:MSRstaticE}
E_{\rm stat}(r,\mu) \,&\,=\, 2\,m_Q^{\rm pole} + V_{\rm QCD}(r,\mu)
=\, 2\,m_Q^{\rm MSR}(R_0) + 2\,\delta m_Q^{\rm MSR}(R_0,\mu)
+ V_{\rm QCD}(r,\mu)\nonumber\\
\,&\,\equiv\,2\,m_Q^{\rm MSR}(R_0) + V^{\rm MSR}_{\rm QCD}(r,\mu,R_0)\,.
\end{align}
However, to avoid the appearance of large logs of $R_0/\mu$ in $\delta m_Q^{\rm MSR}(R_0,\mu)$ we can use \mbox{R-evolution}
to sum up large logs and express it in terms of $\delta m_Q^{\rm MSR}(R,\mu)$\,:
\begin{align}
\delta m_Q^{\rm MSR}(R_0) &= \delta m_Q^{\rm MSR}(R_0) + \delta m_Q^{\rm MSR}(R) - \delta m_Q^{\rm MSR}(R) = \delta m_Q^{\rm MSR}(R) + m_Q^{\rm MSR}(R) - m_Q^{\rm MSR}(R_0) \nonumber\\
& = \delta m_Q^{\rm MSR}(R) + \Delta^{\rm MSR}(R,R_0)\,.
\end{align}
Hence, the following R-improved expression for the MSR-scheme Static QCD Potential can be written as\,:
\begin{align}\label{eq:MSRStat}
V^{\rm MSR}_{\rm QCD}(r,\mu,R_0) =  V_{\rm QCD}(r,\mu) + 2\, \delta^{\rm MSR}(R,\mu) +2\, \Delta^{\rm MSR}(R,R_0)\,.
\end{align}
The R-improved static potential is similar to the Renormalon Subtracted scheme used in Ref.~\cite{Brambilla:2009bi},
based on \cite{Pineda:2001zq}, and to the analysis of Ref.~\cite{Sumino:2003yp} based on renormalon dominance.

The result for Eq.~\eqref{eq:MSRStat} is shown in Figs.~\ref{fig:MSRnbottom} for bottomonium for the value
\mbox{$R_0=1\,$GeV}.\,\footnote{Results for charmonium are similar and can be found in Ref.~\cite{Mateu:2018zym}.} 
The choice of $R_0$ is set so it is identical to the value at which the renormalization scale freezes. 
We can see how the static MSR potential converges nicely towards the Cornell model for moderate values of $r$. The
agreement for large distances improves with each new perturbative order addition.
For larger values of the radius $\log(r\mu)$ becomes large, as $\mu$ freezes, which makes perturbation theory unreliable. At small distances all orders agree very well due to the small value of $\alpha_s$,
but disagree with the Cornell model. In conclusion, the Cornell model and QCD agree for moderate values of
$r$, but disagree in the ultraviolet as the model does not incorporate logarithmic modifications due to the running
of $\alpha_s$. An interesting question is, then, if this difference in the UV can be absorbed in the definition of the quark mass. This confirms the claims of Refs.~\cite{Sumino:2003yp,Sumino:2004ht} where, using renormalon dominance arguments and in the framework of
the operator product expansion of pNRQCD, it is shown that perturbation theory alone should be capable of describing both the Coulomb and linear behavior of the static potential, and that non-perturbative corrections start at
$\mathcal{O}(\LQCD^3r^2)$. 

The aforementioned R-improved QCD static potential can also be compared to lattice simulations. 
Our result for $\sigma$ in Eq.~\eqref{eq:Cornell-Fit} is in very nice agreement with the lattice determinations of Ref.~\cite{Koma:2006fw} ($\sigma = 0.2098 \pm 0.0009\,$GeV$^2$) and 
Ref.~\cite{Kawanai:2013aca} ($\sigma = 0.206 \pm 0.010\,$GeV$^2$). The direct comparison of
$\alpha_s^{\rm Cornell}$ parameter and lattice analyses differs by a factor of roughly $2$.
However, such result should be taken with caution, as in the static potential at short distances loop corrections modify the short-distance $1/r$ behavior. Thus, this discrepancy should be of little concern.

\begin{figure}[tbh!]\centering
\floatbox[{\capbeside\thisfloatsetup{capbesideposition={right,top},capbesidewidth=6cm}}]{figure}[\FBwidth]
{\caption{Comparison of the R-improved perturbative static potential with lattice QCD results for $n_f = 3$ dynamical flavors. We use $\mu=R$, with $\mu$ set to the profile shown in Fig.~\ref{fig:Profile}. We determine
$\alpha_s$ and $R_0$ fitting our pNRQCD theoretical expression to the lattice
simulations~\cite{Bazavov:2014pvz,Bazavov:2017dsy}.}\label{fig:Lattice}}
{\includegraphics[width=\linewidth]{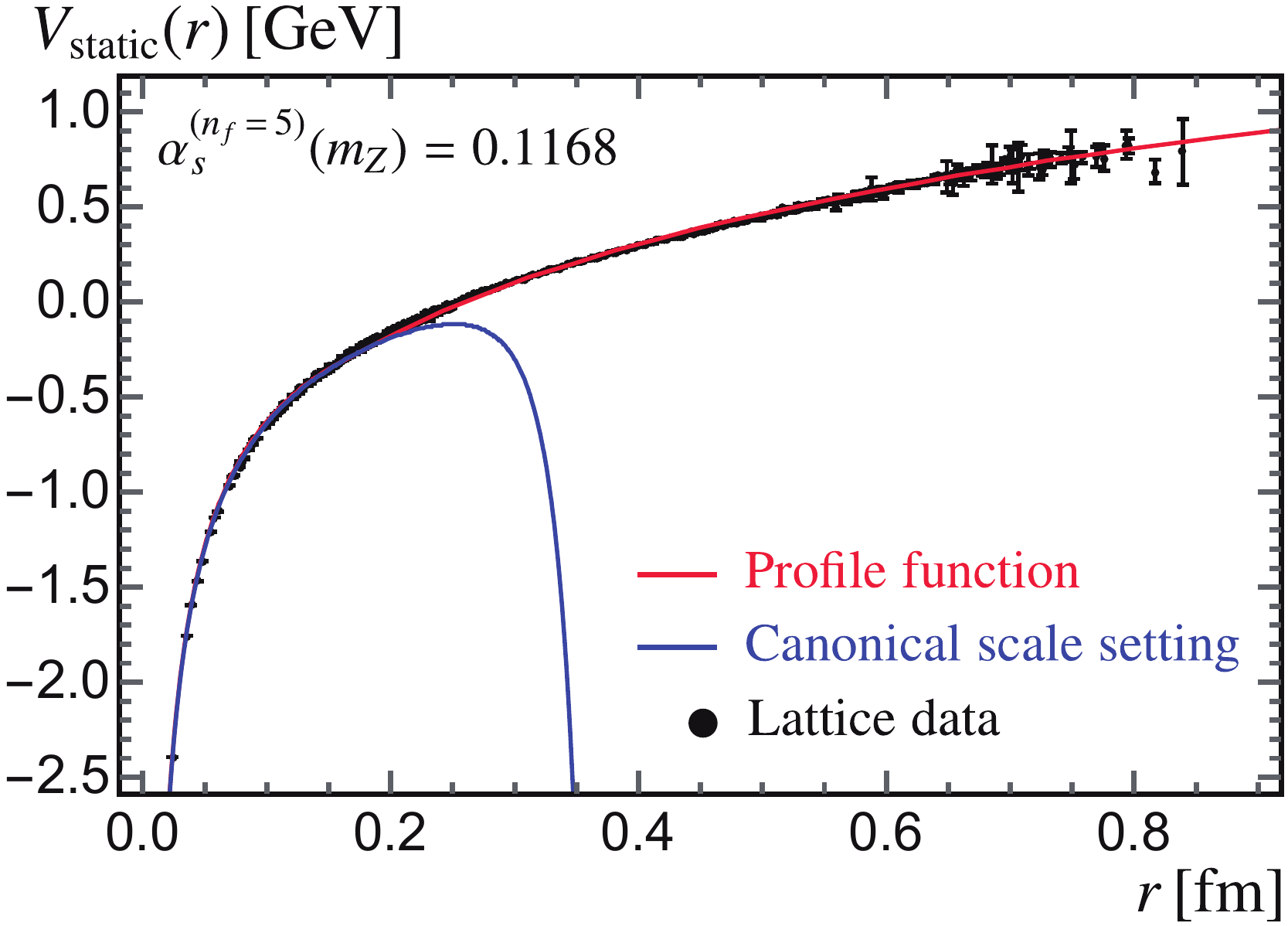}}
\end{figure}

Regarding the static potential, we use the results of Ref~\cite{Bazavov:2014pvz} with lattice
spacing \mbox{$a=0.04\,$fm}. These range between $0.039\,{\rm fm} \leq r \leq 0.84\,$fm, with an average relative precision
of $2.6\,\%$. This dataset is complemented with results from \cite{Bazavov:2017dsy} with a smaller lattice spacing of
$a = 0.025\,$fm, covering values of the radius as small as $0.024\,$fm. For the latter we only consider data with 
\mbox{$r\le0.25\,$fm}, since uncertainties in Ref.~\cite{Bazavov:2017dsy} are larger for higher values of $r$. 
Furthermore, in the range \mbox{$0.024\,{\rm fm}\leq r \leq0.25\,$fm}, Ref.~\cite{Bazavov:2017dsy} 
has more density of points than \cite{Bazavov:2014pvz}.
The full dataset is shown in Fig.~\ref{fig:Lattice} as black dots with error bars. 
The static potential is only dependent on $\alpha_s$ and an arbitrary additive constant, a vertical offset which can be related to the subtraction scale $R_0$ in our R-improved version. For that reason, we perform a two-parameter fit to the lattice data of our \mbox{R-improved} static QCD potential with the scale setting
shown in Fig.~\ref{fig:Profile}. For this simple comparison we use $\mu = R$ and no large ultrasoft logs resummation. The lattice data is assumed to be statistically independent, as the correlation matrix is currently unknown. The results of the fit are\,:
\begin{align}\label{eq:alphaSfit}
\alpha_s^{(n_f = 5)}(m_Z) = 0.1168\,,\qquad R_0 = 1.024\,{\rm GeV}.
\end{align}
Note that the remarkable agreement of the value of $R_0$ to that previously employed to compare with the Cornell model potential.
No fit uncertainties are shown as they would not reflect the actual accuracy reachable by this procedure.
The comparison using the above best-fit values for $\alpha_s$ and $R_0$ is shown as a red
line in Fig.~\ref{fig:Lattice}. The blue line shows the R-improved static QCD potential with the same
values for the parameters, but implementing a fully canonical profile $\mu = 1/r$. Our profile completely agrees with lattice QCD results up to distances of approximately
$1\,$fm, while the canonical profile behaves in an unphysical manner for $r\gtrsim 0.2\,$fm. Such findings can have a direct impact on the precise determination of $\alpha_s$ constant from fits to lattice simulations.

\section{Bottomonium spectrum with a floating bottom quark mass}

\noindent
In this section we will describe the procedure to fit the Cornell potential with NRQCD using synthetic 
spectra with a floating bottomonium mass. This complete bottomonium spectrum, up to $n_p = 2$ for
arbitrary \mbox{$\mbar_b\equiv\mbar_b(\mbar_b)$} is constructed using NRQCD up to N$^3$LO~\cite{Kiyo:2014uca}.
In the pole mass scheme~\cite{Penin:2002zv,Beneke:2005hg,Kiyo:2014uca}, the energy of a non-relativistic $Q\overline{Q}$
bound state, univoquely determined by the $(n_p,j,\ell,s)$ quantum numbers and with $\nl$ massless active flavors, can be written
as\,:\,\footnote{See Ref.~\cite{Mateu:2018zym} for more details.}
\begin{align}\label{eq:EXpole}
& M^{(\nl)}_{n_p,j,\ell,s}(\mu) = 2\,m_Q^{\rm pole}\!
\Bigg[1-\frac{C_F^2\,\alpha^{(\nl)}_s(\mu)^2}{8n_p^2}\sum_{i=0}^{\infty}
\bigg(\frac{\alpha^{(\nl)}_s(\mu)}{4\pi}\bigg)^{\!\!i}\,
\varepsilon^{i+1}P_i(L_{\nl})\Bigg],
\end{align}
with
\begin{align}\label{eq:NRlog}
& L_{\nl}=\log\!\bigg(\frac{n_p\mu}{C_F\alpha_s^{(\nl)}(\mu)m_Q^{\rm pole}}\bigg)+ H_{n_p+\ell}\,,\qquad
P_i(L) = \sum_{j=0}^i\,c_{i,j}\,L^j\,,
\end{align}
where $H_{n}$ is the harmonic number. In Eq.~\eqref{eq:EXpole} $\varepsilon$ acts as a bookkeeping
parameter that properly implements the so called $\Upsilon$-expansion~\cite{Hoang:1998hm}.
The $c_{i,0}$ coefficients have been computed up to $i=3$~\cite{Brambilla:2001qk,Kiyo:2013aea},
while the $c_{i,j>0}$ coefficients can be directly obtained from the latter $c_{i,0}$ imposing $\mu$
independence of the quarkonia mass. We denote the sum in Eq.~\eqref{eq:NRlog} truncated to
$\mathcal{O}(\varepsilon^{n+1})$ as the N$^n$LO result. This formula does not include the resummation of
large ultrasoft logarithms.\,\footnote{See Ref.~\cite{Peset:2018jkf} for a recent summation to N$^3$LL precision for P-wave states.}

Due to the already discussed $u=1/2$ renormalon in the QCD static potential, it is essential to express the static energy in Eq.~\eqref{eq:EXpole} in terms of a short-distance mass, exactly as in the static energy.
Following the results of Sec.~\ref{sec:StaticPot} and the analysis in Ref.~\cite{Mateu:2017hlz} we will employ the MSR mass~\cite{Hoang:2017suc} in our analysis.

The final goal of this study is to obtain relations among the Cornell model constants (especially the Cornell mass) and the QCD fundamental parameters $\alpha_s$ and $m_b$. To achieve this, we will perform a calibration scanning over these two parameters.
Specifically, we will create templates of bottomonium spectra in reasonable ranges of values for the bottom mass and the strong coupling constant, and from there we will obtain the dependence of the Cornell model parameters in terms of $m_b$ and $\alpha_s$. For the former we will vary the $\MSb$ mass between $4$ and $8\,$GeV in steps of $500\,$MeV.
For the strong coupling constant, we consider $\alpha_s^{(n_\ell=5)}(m_Z)$ between $0.114$ and $0.12$ for $\mbar_b = 4.2\,$GeV.

We generate QCD predictions for the $n_p=\{1,2\}$ bottomonium states varying the two renormalization scales $\mu$ and $R$ in a correlated way (see Ref~\cite{Mateu:2018zym} for details). The ranges are selected so the argument of the logarithm in Eq.~\eqref{eq:NRlog} in the MSR scheme ranges between $1/2$ and $2$. Indeed, this range in general depends on the value of $m_b$. Hence, we set a lower limit dependent on the bottom mass: 
\begin{align}\label{eq:mui}
\mu_1^{\rm min} = 0.638\,{\rm GeV} + 0.209\,\overline{m}_b\,\,,\,\,\,\,
\mu_2^{\rm min} = 0.510\,{\rm GeV} + 0.120\,\overline{m}_b\,,
\end{align}
where the subindex $1,\,2$ denotes the principal quantum number of the heavy quarkonium state.
The upper limit is fixed to $4\,$GeV, as no dependence of $m_b$ is found for $\mu$ or $R$. 

Given the synthetic bottomonium spectrum for a specific $m_b$ and $\alpha_s$, we will fit the Cornell model parameters via a simple statistical regression analysis. The generated QCD pseudo-data at LO, NLO, N$^2$LO and
N$^3$LO can be thought of as a set of highly correlated experimental measurements, which makes a traditional $\chi^2$
fit with a non-diagonal covariance matrix impossible, due to the d'Agostini bias. However, the theoretical
covariance matrix is the only uncertainty source we have, so it is impossible to write down a $\chi^2$ function. 
We follow the strategy of Ref.~\cite{Mateu:2017hlz}, where the QCD renormalization scales were varied in a correlated way in terms of two dimensionless variables \mbox{$\mu_n = \mu_n^{\rm min} + x\,(4\,{\rm GeV} - \mu_n^{\rm min})$},
\mbox{$R_n = \mu_n^{\rm min} + y\,(4\,{\rm GeV} - \mu_n^{\rm min})$}, with \mbox{$n = \{1, 2\}$, $0\le \{x,y\}\le1$} and
$\mu_i$ defined in Eq.~\eqref{eq:mui}. The employed regression $\chi^2$ function is, then\,:
\begin{align}\label{eq:chi2Reg}
\chi^2(x,y,\mbar_b)=\frac{\sum_i\big(M_i^{\rm QCD}(x,y,\mbar_b)-M_i^{\rm Cornell}\big)^{2}}
{\frac{\chi^2_{\rm min}(x,y)}{\rm d.o.f.}}\,,
\end{align}
where, actually, $\chi^2_{\rm min}(x,y)$ is known only after the minimization is carried out, but allows the $\chi^2$
function to be dimensionless. For a given value of $\mbar_b$, all possible
values of $\{x,y\}$ are scanned, and the parameters of the Cornell model are determined for each pair. Scanning over all values of $\mbar_b$ and $\alpha_s$, functions of the Cornell model
parameters in terms of these QCD fundamental quantities are obtained.

\section{Calibration of Cornell model}\label{sec:results}
\begin{figure}[tbh!]\centering
\floatbox[{\capbeside\thisfloatsetup{capbesideposition={right,top},capbesidewidth=6cm}}]{figure}[\FBwidth]
{\caption{Upper plot\,: Dependence of the Cornell mass parameter with the MSR bottom mass at
$R=1\,$GeV at N$^3$LO, using the world average value for the strong coupling constant. \newline
Lower plot\,: Difference between the
Cornell and MSR masses as a function of $m_b^{\rm MSR}(R=1\,{\rm  GeV})$. \newline
The error bars represent the quadratic sum of fit and perturbative
uncertainties. The blue line and band in the upper plot shows a linear fit to the points, where the individual uncertainties are taken as uncorrelated. In the lower plot, those correspond to the weigthed average of the differences and the regular average of the uncertainties, respectively.}
\label{fig:Money}}
{\includegraphics[width=0.92\linewidth]{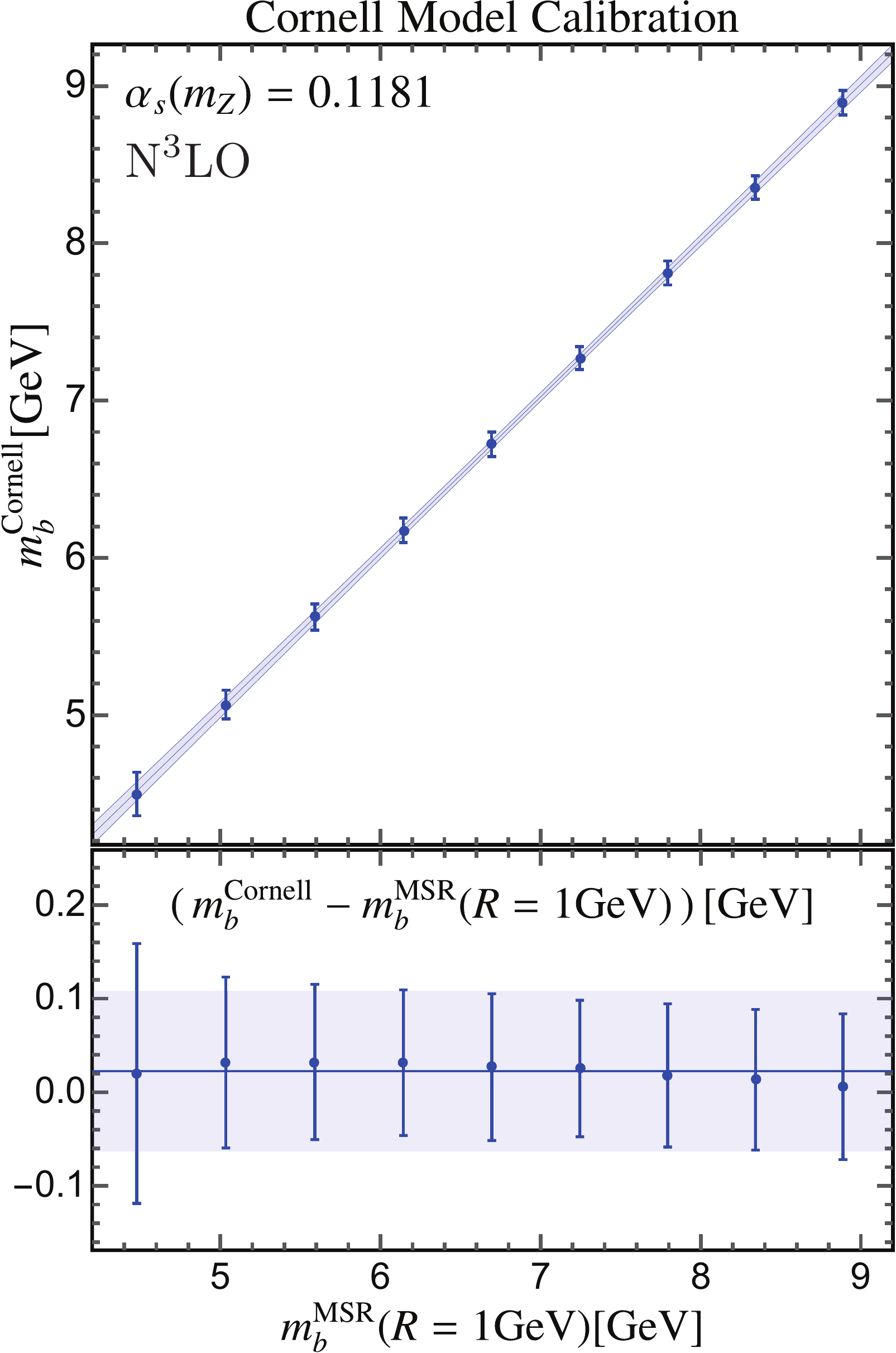}}
\end{figure}

\noindent
First we focus on the dependence of the Cornell mass parameter with the short-distance QCD mass.
For convenience, we have produced our bottomonium spectra as a function of $\MSb$ mass, although this scheme should be thought as a coupling constant rather than a kinematic mass.
Following the analysis of Secs.~\ref{sec:StaticPot}, we use the MSR mass with small values of $R$, which is a kinetic mass free from renormalon ambiguities. When we calibrate the Cornell mass versus 
\mbox{$m_b^{\rm MSR}(R={1\rm GeV})$} we observe a linear relation, with a slope very close to (and compatible with) unity ($0.995\pm 0.026$) and a constant term compatible with zero ($0.05 \pm 0.19$). 
Furthermore, this correspondance is achieved for all considered values of $\alpha_s$ and different orders~\cite{Mateu:2018zym}.
In Fig.~\ref{fig:Money} we show the linear relation at N$^3$LO for the world average value of $\alpha_s$, even though this correspondance is achieved for all considered values of $\alpha_s$ and different perturbative orders. 

The most relevant result of this work is, however, the difference between the Cornell model mass parameter and the MSR mass\,:
\begin{align}\label{eq:main-result}
m_b^{\rm Cornell} = m_b^{\rm MSR}(R = 1\,{\rm GeV}) + [\,0.023 \pm 0.086\,{\rm GeV}\,]\,,
\end{align}
calculated as the weigthed average of the individual
$m_b^{\rm Cornell} - m_b^{\rm MSR}(1\,{\rm GeV})$ values, taking the uncertainty as the regular average of individual uncertainties.

\begin{figure*}[tbh!]
\center
{\includegraphics[width=0.46\textwidth]{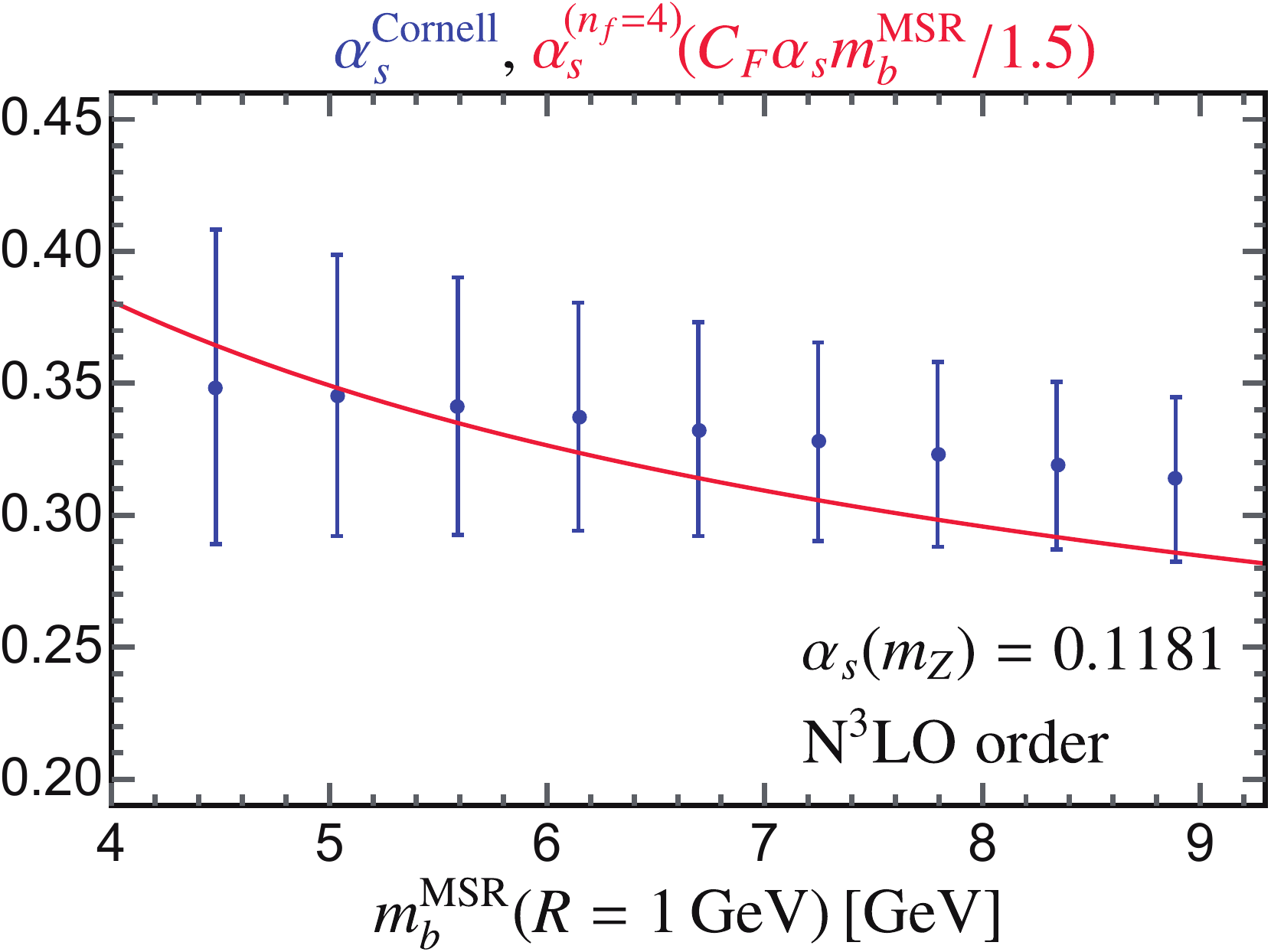}~~~~}
{\includegraphics[width=0.47\textwidth]{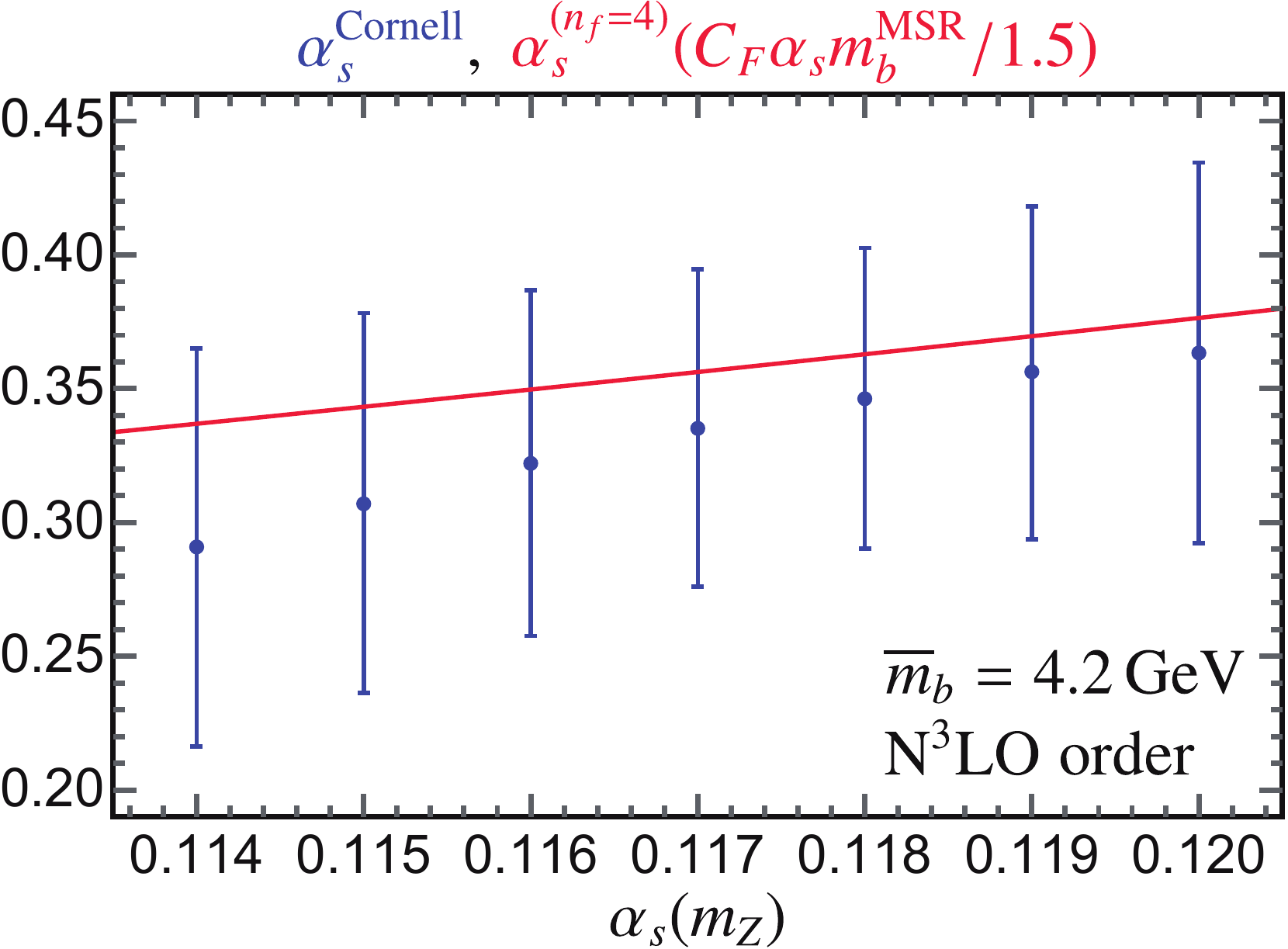} }\vspace*{-0.3cm}
{\includegraphics[width=0.46\textwidth]{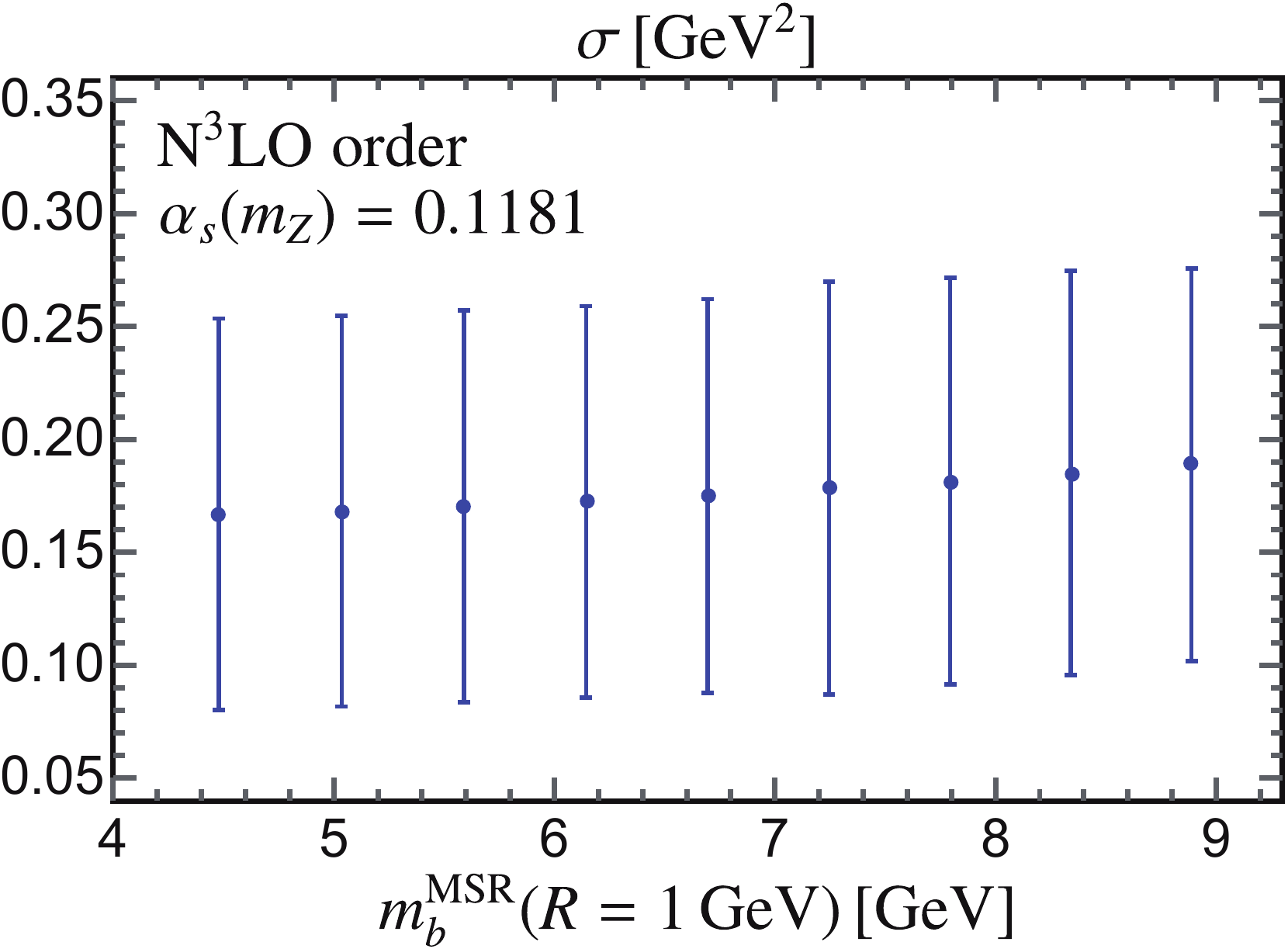}~~~~}
{\includegraphics[width=0.47\textwidth]{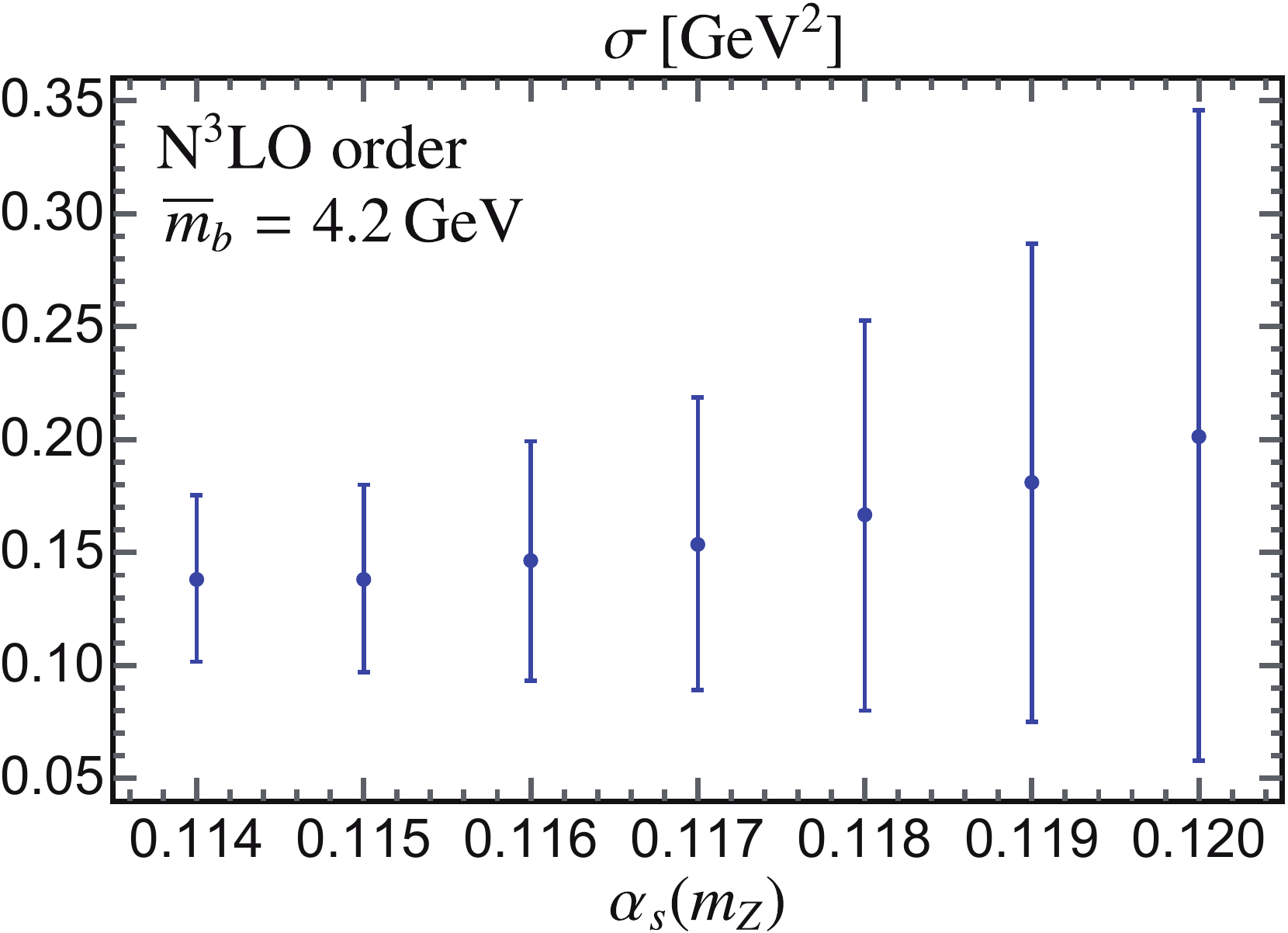} }
\caption{Dependence of the Cornell parameters $\alpha_s^{\rm Cornell}$ (upper two panels)
and $\sigma$ (lower two panels) with the MSR bottom mass (leftmost two panels) and the QCD coupling constant at the
Z-pole (rightmost two panels). The two upper plots also show, with a solid red line, the strong coupling constant evaluated at a characteristic non-relativistic scale.}
\label{fig:alpha-sigma}
\end{figure*}

Finally, we show the dependence of the remaining Cornell parameters, $\alpha_s^{\rm Cornell}$
and  $\sigma$, with the QCD quantities $\alpha_s$ and $m_b^{\rm MSR}$.  
The variation of $\alpha_s^{\rm Cornell}$ with the latter QCD parameters is shown in the upper plots of Fig.~\ref{fig:alpha-sigma} as blue dots. The red solid line corresponds to the QCD strong coupling constant evaluated at the non-relativistic scale $\mu_{\rm NR}$, chosen so the argument of the log in Eq.~\eqref{eq:NRlog} is of the order of one. As our fit employs $n_p=\{1,2\}$ states, the average value $\bar n_p=1.5$ is taken.

The scale $\mu_{\rm NR}$ is obtained by choosing $R=\mu$ and solving numerically the equation\,:
\begin{align}\label{eq:muNR}
C_F\alpha_s^{(n_f=4)}(\mu_{\rm NR})\,m_b^{\rm MSR}(\mu_{\rm NR}) = \frac{3}{2}\,\mu_{\rm NR}\,.
\end{align}
This scale is determined for different values of $\alpha_s^{(n_f=5)}(m_Z)$ and the bottom mass, finding a 
remarkable agreement between $\alpha_s^{\rm Cornell}$ and $\alpha_s^{(n_f=4)}(\mu_{\rm NR})$ within perturbative uncertainties. Besides the similar order of magnitude, its dependence on $m_b$ and $\alpha_s(m_Z)$ follows the same trend. This simple analysis disfavors quark models which consider a QCD-like running for $\alpha_s^{\rm Cornell}$, with a scale given by the reduced mass of the $Q\overline Q$ pair.

The lower plots in Fig.~\ref{fig:alpha-sigma} represents the dependence of $\sigma$ with the bottom mass and the strong coupling constant. It shows a remarkable independence on the bottom mass, as expected for a static potential. Taking the average of all dots we obtain \mbox{$\sigma^{\rm Cal.} = 0.176 \pm 0.088\,$GeV$^2$}, which compares well with that obtained from the fit to the bottomonium experimental data in
Eq.~\eqref{eq:Cornell-Fit} \mbox{$\sigma^{\rm fit} = 0.207 \pm 0.011\,{\rm GeV}^2$}. Regarding the dependence on  $\alpha_s^{(n_f=5)}(m_Z)$, a small positive correlation is envisaged, although given the uncertainties no strong conclusions can be drawn. Anyway, some dependence of $\sigma$ with $\alpha_s$ is expected since, as argued in Refs.~\cite{Sumino:2003yp,Sumino:2004ht} and confirmed in Sec.~\ref{sec:StaticPot} of this work, the linear rising term in the static potential is of perturbative nature.

\bibliographystyle{JHEP}
\bibliography{NRQCD}

\end{document}